\documentstyle[aps]{revtex}

\newcommand{\bea}{\begin{eqnarray}}
\newcommand{\eea}{\end{eqnarray}}

\begin{document}

\draft
%%%%%%%%%%%%%%%%%%%%%%%%%%%%%%%%%%%%%%%%%%%%%%%%%%%%%%%%%%%%%%%
\title{Cosmological Perturbations in Generalized Gravity Theories: \\
       Conformal Transformation}
\author{Jai-chan Hwang}
\address{Department of Astronomy and Atmospheric Sciences,
         Kyungpook National University, Taegu, Korea}
\date{\today} 
\maketitle

%%%%%%%%%%%%%%%%%%%%%%%%%%%%%%%%%%%%%%%%%%%%%%%%%%%%%%%%%%%%%%%
\begin{abstract}

A broad class of generalized Einstein's gravity can be cast into
Einstein's gravity with a minimally coupled scalar field using 
suitable conformal rescaling of the metric.  Using this conformal 
equivalence between the theories, we derive the equations for the 
background and the perturbations, and the general asymptotic solutions
for the perturbations in the generalized Einstein's gravity
from the simple results known in the minimally coupled scalar field.
Results for the scalar and tensor perturbations can be presented in 
unified forms.  The large scale evolutions for both modes are 
characterized by corresponding conserved quantities.  We also present 
the normalization condition for canonical quantization.

\end{abstract}


\widetext
%%%%%%%%%%%%%%%%%%%%%%%%%%%%%%%%%%%%%%%%%%%%%%%%%%%%%%%%%%%%%%%
\section{Introduction}
         \label{sec:Introduction}

Studies of generalized forms of Einstein's 
theory as the theories for the gravity have been made by many authors.
Some of the widely studied generalized gravity theories include
Brans-Dicke theory, induced gravity, dilaton coupling, nonminimally
coupled scalar field, nonlinear scalar curvature coupling, etc.
We have variety of motivations for considering more generalized forms 
for the gravity: Mach principle, quantum backreaction, renormalization,
higher dimensional unification, cosmology, etc.
In the context of generalized gravity theories, the Einstein's gravity can 
be regarded as a limiting case.

In \cite{H-GGT} we presented a thorough derivation of 
the equations and general asymptotic solutions describing the
scalar and tensor type perturbations in the conventional cosmological
spacetime supported by the generalized gravity.
The progress made in \cite{H-GGT} can be summarized as finding the role
of a proper choice of the gauge in treating the scalar mode;
previous studies in \cite{H-GGT1}-\cite{H-GGT3} were made in 
a different gauge condition. 
We found that by employing a suitable gauge the scalar mode can be 
described in a simple manner similarly as in the Einstein's gravity.
When we deal with the minimally coupled scalar field we found that the
choice of uniform-curvature gauge, or equivalently corresponding
gauge invariant combination of variables, simplifies the problem;
see \cite{H-QFT}-\cite{H-IF}.
In \cite{H-GGT} we also found that when we deal with the generalized
gravity involving the scalar field and the scalar curvature, the
uniform-curvature gauge again suits the problem.
In fact, we discovered that in the large scale limit, 
neglecting the transient mode, the same solution for the minimally
coupled scalar field remains valid in a broad class of
generalized Einstein's gravity.
The equations and the general solutions for both the scalar and tensor 
modes can be presented in unified forms.

It is known in the literature that using a conformal transformation
the class of generalized gravity theories we mentioned can be cast into 
Einstein's gravity with a minimally coupled scalar field,
\cite{Dicke,H-GGT1}.
Using the mathematical equivalence between the generalized gravity
theories and the simple minimally coupled scalar field 
under the conformal transformation,
without losing any rigour, we can derive the equations and solutions 
in the generalized gravity directly from the known results in the 
Einstein's gravity.
In this paper we use the conformal equivalence between the theories
only as a mathematical tool; for some discussions concerning
the physics, see \cite{Brans}.
The conformal transformation properties of the background and perturbed 
quantities in the cosmological spacetime are presented in 
\cite{H-GGT1,H-GGT3}. 
However, at the time of the work in \cite{H-GGT1}-\cite{H-GGT3}
the proper role of the uniform-curvature gauge was not known.
Using the conformal transformation the rigorously derived results 
in \cite{H-GGT} can be rederived in a considerably simple manner. 
In this paper we will present the derivation.
We will also summarize the results for the individual gravity case
in tabular forms.

\vskip 1cm
%%%%%%%%%%%%%%%%%%%%%%%%%%%%%%%%%%%%%%%%%%%%%%%%%%%%%%%%%%%%%%
\centerline{\bf II. GENERALIZED $f(\phi, R)$ GRAVITY}
\stepcounter{section}
\vskip .5cm

We consider a general class of gravity theories with the Lagrangian
\bea
   & & L = {1\over 2} f(\phi, R) 
       - {1 \over 2} \omega(\phi) \phi^{;a} \phi_{,a} - V(\phi).
   \label{GGT-Lagrangian}
\eea
The gravitational field equation and the equation of motion 
for the scalar field are:
\bea
   & & G_{ab} = {1\over F} \left[ \omega \left( \phi_{,a} \phi_{,b}
       - {1\over 2} g_{ab} \phi^{;c} \phi_{,c} \right)
       - g_{ab} {RF - f + 2 V \over 2} + F_{,a;b} - g_{ab} {F^{;c}}_{c} 
       \right],
   \label{GGT-GFE} \\
   & & {\phi^{;a}}_{a} + {1 \over 2 \omega} \left(
       \omega_{,\phi} \phi^{;a} \phi_{,a}
       + f_{,\phi} - 2 V_{,\phi} \right) = 0.
   \label{GGT-EOM}
\eea
where we defined $F \equiv \partial f / (\partial R)$.
We call it the generalized $f(\phi,R)$ gravity theory.
It includes diverse classes of gravity theories as cases; see Table 1.

\centerline{\bf LOCATE TABLE 1.}

%%%%%%%%%%%%%%%%%%%%%%%%%%%%%%%%%%%%%%%%%%%%%%%%%%%%%%%%%%%%%%%
\section{Conformal transformation to Einstein's gravity}
         \label{sec:CT}

Using the conformal transformation the generalized $f(\phi,R)$ gravity 
can be transformed to the Einstein gravity with an additional scalar field.
By the conformal transformation the metric is redefined as
\bea
   & & \hat g_{ab} = \Omega^2 g_{ab},
   \label{CT}
\eea
where $\Omega$ is a spacetime position dependent factor.
We use a hat to denote quantities based on conformally transformed metric frame.
By defining the conformal factor as
\bea
   & & \Omega \equiv \sqrt{F} \equiv e^{{1 \over 2} \sqrt{2\over 3} \psi}, 
   \label{CT-factor}
\eea
where $\psi$ is a new dynamical variable, one can show that 
(\ref{GGT-Lagrangian}) can be transformed into 
[for derivation, see (42,43) of \cite{H-GGT1}]
\bea
   & & \hat L = {1\over 2} \hat R 
       - {\omega \over F} {1\over 2} \phi^{\hat ; a} \phi_{\hat , a} 
       - {1 \over 2} \psi^{\hat ; a} \psi_{\hat , a} - \hat V ( \phi,\psi), 
       \quad \hat V (\phi,\psi) \equiv {R F - f + 2 V \over 2 F^2}. 
   \label{CT-Lagrangian}
\eea
Thus, our original generalized $f(\phi, R)$ gravity is cast into 
the Einstein theory with an additional scalar field, $\psi$, 
and a special potential term $\hat V(\phi,\psi)$.
In general, $\phi$ and $\psi$ are dependent on each other.
{\it Assuming} $\psi = \psi (\phi)$ and introducing a new scalar field
$\hat \phi$ with $\hat \phi = \hat \phi (\phi)$, (\ref{CT-Lagrangian})
can be transformed into a Lagrangian for the minimally coupled scalar field 
$\hat \phi$
\bea
   & & \hat L = {1\over 2} \hat R - {1 \over 2} 
       \hat \phi^{\hat ; a} \hat \phi_{\hat , a} - \hat V (\hat \phi).
   \label{CT-MSF-Lagrangian}
\eea
{}For this, $\hat \phi$ should satisfy
\bea
   & & d \hat \phi = \sqrt{ {\omega \over F} d \phi^2 + d \psi^2 }.
   \label{hat-phi-phi}
\eea

%%%%%%%%%%%%%%%%%%%%%%%%%%%%%%%%%%%%%%%%%%%%%%%%%%%%%%%%%%%%%%%
\section{Perturbed universe model}
         \label{sec:Model}

As the metric describing the model universe, we consider a spatially 
homogeneous, isotropic, and flat (FLRW) background and general 
perturbations of the scalar and tensor type 
\bea
   & & d s^2 = - \left( 1 + 2 \alpha \right) d t^2
       - a \beta_{,\alpha} d t d x^\alpha
       + a^2 \left[ g_{\alpha\beta}^{(3)} \left( 1 + 2 \varphi \right)
       + 2 \gamma_{|\alpha\beta} + 2 H_T Y^{(t)}_{\alpha\beta}
       \right] d x^\alpha d x^\beta.
   \label{metric-general}
\eea
$a(t)$ is the cosmic scale factor.
$g^{(3)}_{\alpha\beta}$ is a comoving part of the background
three-space metric and a vertical bar indicates a covariant
derivative based on $g^{(3)}_{\alpha\beta}$;
in the flat FLRW background we have 
$g_{\alpha\beta}^{(3)} = \delta_{\alpha\beta}$.
$Y^{(t)}_{\alpha\beta} ({\bf x})$ is a symmetric, tracefree,
and transverse harmonic function with
$\nabla^2 Y^{(t)}_{\alpha\beta} = - k^2 Y^{(t)}_{\alpha\beta}$;
$\nabla^2$ is a Laplacian operator based on $g_{\alpha\beta}^{(3)}$.
The perturbative order quantities $\alpha ({\bf x}, t)$, 
$\beta ({\bf x}, t)$, $\gamma ({\bf x}, t)$, and $\varphi ({\bf x}, t)$ 
characterize the scalar mode, whereas $H_T ({\bf x}, t)$
characterizes the tensor mode.
$\beta$ and $\gamma$ are affected by the spatial coordinate 
transformation in the FLRW spacetime. 
Since the FLRW spacetime is spatially homogeneous and isotropic
we can easily avoid using these spatially gauge dependent variables.
A combination $\chi ({\bf x}, t) \equiv a ( \beta + a \dot \gamma)$ is such
a variable that is spatially gauge invariant.
For the scalar field we let
$\phi ({\bf x}, t) = \bar \phi (t) + \delta \phi ({\bf x}, t)$
where an overbar indicates the background quantity; 
we neglect the overbar unless it is necessary.
Now, the variables $\alpha$, $\varphi$, $\chi$, and $\delta \phi$ 
are spatially gauge invariant, but are temporally gauge dependent.
$H_T$ is gauge invariant.
{}For the gauge transformation properties, see \S 2.2 of \cite{H-PRW}.

We decompose the conformal factor $\Omega$ into the background and 
the perturbed part as 
\bea
   & & \Omega ({\bf x}, t) \equiv \bar \Omega (t) 
       \Big[ 1 + \delta \Omega ({\bf x}, t) \Big].
   \label{Omega-pert}
\eea
{}From (\ref{CT-factor}), we have:
\bea
   & & \bar \Omega = \sqrt{F} = e^{{1\over 2} \sqrt{{2\over 3}} \psi}, 
       \quad \delta \Omega = {\delta F \over 2 F} 
       = {1\over 2} \sqrt{2\over 3} \delta \psi.
   \label{Omega-psi}
\eea
We can show that the only changes under the conformal transformation
are the following [see \S 4.1 of \cite{H-GGT1} and \S 2.1 of \cite{H-PRW}]: 
\bea
   & & \hat a = a \bar \Omega, \quad d \hat t = \bar \Omega d t, \quad
       \hat \alpha = \alpha + \delta \Omega, \quad 
       \hat \varphi = \varphi + \delta \Omega. 
   \label{CT-pert}
\eea
Thus, for example, we have ($H \equiv \dot a / a$):
\bea
   & & \hat H = {1\over \Omega} \left( H + {\dot \Omega \over \Omega} \right),
       \quad \hat \chi = \Omega \chi.
\eea
{}From (\ref{hat-phi-phi}) we can show:
\bea
   & & \dot {\hat \phi} = \sqrt{ {\omega \over F} \dot \phi^2
       + {3 \dot F^2 \over 2 F^2} }, \quad
       {\delta \hat \phi \over \dot {\hat \phi}}
       = {\delta \phi \over \dot \phi}
       = {\delta F \over \dot F}.
   \label{hat-phi-phi-pert}
\eea
Relations among $\hat \phi$, $\phi$, and $F$ in the individual
gravity cases are summarized in Table 2.

\centerline{\bf LOCATE TABLE 2.}

%%%%%%%%%%%%%%%%%%%%%%%%%%%%%%%%%%%%%%%%%%%%%%%%%%%%%%%%%%%%%%%
\section{Perturbations in the Einstein's gravity}
         \label{sec:MSF}

We consider a minimally coupled scalar field.
The Lagrangian is given in (\ref{CT-MSF-Lagrangian}).
In this section we {\it neglect} hats on the background and perturbed
quantities.
The equations describing the evolution of the background
are [see (2-4) of \cite{H-QFT}]:
\bea
   & & H^2 = {1 \over 3} \left( {1 \over 2} \dot \phi^2 + V \right), \quad
       \dot H = - {1 \over 2} \dot \phi^2, \quad
       \ddot \phi + 3 H \dot \phi + V_{,\phi} = 0.
                          \label{BG-MSF}
\eea
The third equation follows from first two.

When we manage the gravity theory involving the scalar field
our experience tells that the uniform-curvature gauge is
the most convenient; equivalently we can take the gauge invariant
variables with a subindex $\varphi$.
{}For the uniform-curvature gauge, see \S 3 of \cite{H-UCG}.
Thorough perturbation analyses for a minimally coupled scalar field
in the uniform-curvature gauge were made in \cite{H-QFT}-\cite{H-IF}.
An example of the gauge invariant combination is
\bea
   & & \delta \phi_\varphi \equiv \delta \phi - {\dot \phi \over H} \varphi
       \equiv - {\dot \phi \over H} \varphi_{\delta \phi}.
   \label{UCG-UFG}
\eea
$\delta \phi_\varphi$ becomes $\delta \phi$ in the uniform-curvature
gauge which takes $\varphi \equiv 0$ as the gauge condition.
The action expanded to the second order in perturbed scalar field is 
presented in (39) of \cite{H-UCG} as
\bea
   & & S = {1\over 2} \int a^3 \Bigg\{ \delta \dot \phi_\varphi^2
       - {1\over a^2} \delta \phi_\varphi^{\;\; |\alpha} 
       \delta \phi_{\varphi,\alpha}
       + {H \over a^3 \dot \phi} \left[ a^3 \left( {\dot \phi \over H}
       \right)^\cdot \right]^\cdot \delta \phi_\varphi^2 \Bigg\} 
       \sqrt{ g^{(3)} } dt d^3 x,
   \label{Action-MSF}
\eea
where in the flat FLRW background we have $g^{(3)} = 1$.
A closed form equation for the {\it scalar field} perturbation 
and the large and small scale asymptotic solutions are
[see (7,22,12,16) of \cite{H-QFT}]:
\bea
   & & \delta \ddot \phi_\varphi + 3 H \delta \dot \phi_\varphi 
       - \left\{ {1 \over a^2} \nabla^2 
       + {H \over a^3 \dot \phi} \left[ a^3 \left( {\dot \phi \over H} 
       \right)^\cdot \right]^\cdot \right\} \delta \phi_\varphi = 0,
   \label{MSF-delta-phi-eq} \\
   & & \delta \phi_\varphi ({\bf x},t)
       = {\dot \phi \over H} \left[ - C ({\bf x})
       + D ({\bf x}) \int^t {H^2 \over a^3 \dot \phi^2 } dt \right],
   \label{MSF-delta-phi-LS-sol} \\
   & & \delta \phi_\varphi ({\bf k}, \eta) = {1 \over a}
       \Big[ c_1 ({\bf k}) e^{i k \eta} + c_2 ({\bf k}) e^{-i k \eta} \Big],
   \label{MSF-delta-phi-SS-sol}
\eea
where we used $d \eta \equiv dt/a$ and $\nabla^2 \rightarrow -k^2$.
$C({\bf x})$ and $D({\bf x})$ in the large scale solution are 
the coefficients of the growing and decaying modes, respectively.
Quantum field theoretical analyses of (\ref{MSF-delta-phi-eq}) 
in the context of cosmological curved spacetime can be found in 
\cite{H-QFT,H-UCG,H-IF}.
Solutions in (\ref{MSF-delta-phi-LS-sol},\ref{MSF-delta-phi-SS-sol})
are valid for a general $V(\phi)$.
Using (\ref{BG-MSF}), (\ref{MSF-delta-phi-eq}) can be written in 
a compact form as [see (11) of \cite{H-QFT}]
\bea
   & & {H \over a^3 \dot \phi} \left[ {a^3 \dot \phi^2 \over H^2}
       \left( {H \over \dot \phi} \delta \phi_\varphi \right)^\cdot 
       \right]^\cdot - {1 \over a^2} \nabla^2 \delta \phi_\varphi = 0.
   \label{MSF-delta-phi-eq-compact}
\eea

The equation and the asymptotic solutions for the {\it gravitational wave} are
[see (101) of \cite{H-PRW}]:
\bea
   & &  \ddot H_T + 3 H \dot H_T - {1 \over a^2} \nabla^2 H_T = 0,
   \label{MSF-GW-eq} \\
   & & H_T ({\bf x}, t) = C_g ({\bf x})
       - D_g ({\bf x}) \int^t_0 {1 \over a^3} dt,
   \label{MSF-GW-LS-sol} \\
   & & H_T ({\bf k}, \eta) = - { 1 \over a }
       \Big[ c_{g1} ({\bf k}) e^{ik\eta} + c_{g2} ({\bf k}) e^{-ik\eta} \Big].
   \label{MSF-GW-SS-sol}
\eea
The action for the gravitational wave mode can be found in 
\S 18 of \cite{Mukhanov-etal}.
The {\it vorticity mode} does not directly couple with the scalar type 
gravity theory; it will evolve according to the angular momentum conservation
in the expanding background [see \S 3.2.2 of \cite{H-GGT1}].

%%%%%%%%%%%%%%%%%%%%%%%%%%%%%%%%%%%%%%%%%%%%%%%%%%%%%%%%%%%%%%%
\section{Perturbations in generalized Einstein's gravities}
         \label{sec:GGT}

Using (\ref{CT-pert}) we can show that the following quantities
are invariant under the conformal transformation:
\bea
   & & d \eta, \quad \nabla^2, \quad k, \quad \varphi_{\delta \phi}, \quad 
       {H \over \dot \phi} \delta \phi_\varphi, \quad H_T.
   \label{CT-invariants}
\eea
We regard the quantities in \S \ref{sec:MSF}, 
(\ref{BG-MSF}-\ref{MSF-GW-SS-sol}), 
are in the conformal frame, thus with hats on them.
Using the conformal transformation properties in 
(\ref{Omega-pert}-\ref{hat-phi-phi-pert},\ref{CT-invariants}) we can 
derive the corresponding counterparts in the original frame which are now 
valid for gravity theories included in the generalized $f(\phi,R)$ gravity.

{}For the background, (\ref{BG-MSF}) leads to
\bea
   & & H^2 = {1 \over 3F} \left( {\omega \over 2} \dot \phi^2
       + {RF - f + 2 V \over 2} - 3 H \dot F \right),
                          \label{BG-GGT1} \\
   & & \dot H = - {1\over 2 F} \left( \omega \dot \phi^2
       + \ddot F - H \dot F \right),
                          \label{BG-GGT2} \\
   & & \ddot \phi + 3 H \dot \phi + {1\over 2 \omega}
       \left( \omega_{,\phi} \dot \phi^2 - f_{,\phi} + 2 V_{,\phi}
       \right) = 0.
                          \label{BG-GGT3}
\eea
When we derive (\ref{BG-GGT3}) we may need $R = 6 ( 2 H^2 + \dot H)$;
(\ref{BG-GGT3}) also follows from (\ref{BG-GGT1},\ref{BG-GGT2}).

The same form of (\ref{UCG-UFG}) remains valid in the generalized Einstein's 
gravity.
The action in (\ref{Action-MSF}) becomes (we ignore the surface terms)
\bea
   S 
   &=& {1\over 2} \int a^3 
       \left. { \omega + {3 \dot F^2 \over 2 \dot \phi^2 F} 
       \over \left( 1 + {\dot F \over 2 H F} \right)^2 }
       \right\{ \delta \dot \phi_\varphi^2
       - {1 \over a^2} \delta \phi_\varphi^{\;\; |\alpha} 
       \delta \phi_{\varphi,\alpha}
   \nonumber \\
   & & 
       \left. + {H \over a^3 \dot \phi} 
       { \left( 1 + {\dot F \over 2 H F} \right)^2 
       \over  \omega + {3 \dot F^2 \over 2 \dot \phi^2 F} }
       \left[ a^3 
       { \omega + {3 \dot F^2 \over 2 \dot \phi^2 F} 
       \over \left( 1 + {\dot F \over 2 H F} \right)^2 }
       \left( {\dot \phi \over H} \right)^\cdot \right]^\cdot 
       \delta \phi_\varphi^2 \right\}
       \sqrt{ g^{(3)} } dt d^3 x.
   \label{Action-GGT}
\eea
{}For the scalar mode, (\ref{MSF-delta-phi-eq}-\ref{MSF-delta-phi-SS-sol}) 
lead to:
\bea
   & & \delta \ddot \phi_\varphi
       + \left\{ 3 H + { \left( 1 + {\dot F \over 2 H F} \right)^2 \over
       \omega + {3 \dot F^2 \over 2 \dot \phi^2 F } }
       \left[ { \omega + {3 \dot F^2 \over 2 \dot \phi^2 F } \over
       \left( 1 + {\dot F \over 2 H F} \right)^2 } \right]^\cdot \right\}
       \delta \dot \phi_\varphi
   \nonumber \\
   & & \qquad
       - \left\{ {1 \over a^2} \nabla^2 
       + {H \over a^3 \dot \phi} { \left( 1 + {\dot F \over 2 H F} \right)^2 
       \over \omega + {3 \dot F^2 \over 2 \dot \phi^2 F} }
       \left[ { \omega + {3 \dot F^2 \over 2 \dot \phi^2 F} \over
       \left( 1 + {\dot F \over 2 H F} \right)^2 }
       a^3 \left( {\dot \phi \over H} \right)^\cdot \right]^\cdot
       \right\} \delta \phi_\varphi = 0,
   \label{GGT-delta-phi-eq} \\
   & & \delta \phi_\varphi ({\bf x}, t)
       = {\dot \phi \over H}
       \left[ - C ({\bf x}) + D ({\bf x}) \int_0^t
       { \left( H + {\dot F \over 2 F} \right)^2 \over
       a^3 \left( \omega \dot \phi^2 + {3 \dot F^2 \over 2 F} \right) }
       dt \right],
   \label{GGT-delta-phi-LS-sol} \\
   & & \delta \phi_\varphi ({\bf k}, \eta)
       = {1 \over a} { 1 + {\dot F \over 2 H F} \over
       \sqrt{ \omega + {3 \dot F^2 \over 2 \dot \phi^2 F} } }
       \Big[ c_1 ({\bf k}) e^{i k \eta} + c_2 ({\bf k}) e^{i k \eta} \Big].
   \label{GGT-delta-phi-SS-sol}
\eea
In order to derive (\ref{GGT-delta-phi-eq}) it is much simpler to use
(\ref{MSF-delta-phi-eq-compact}) which leads to
\bea
   & & {\dot \phi \over H} 
       { \left( H + {\dot F \over 2 F} \right)^2 \over
       a^3 \left( \omega \dot \phi^2 + {3 \dot F^2 \over 2 F} \right) }
       \left[ { a^3 \left( \omega \dot \phi^2 + 
       {3 \dot F^2 \over 2 F} \right) \over
       \left( H + {\dot F \over 2 F} \right)^2 }
       \left( {H \over \dot \phi} \delta \phi_\varphi \right)^\cdot 
       \right]^\cdot
       - {1 \over a^2} \nabla^2 \delta \phi_\varphi = 0.
   \label{GGT-delta-phi-eq-compact}
\eea
By expanding (\ref{GGT-delta-phi-eq-compact}) we get (\ref{GGT-delta-phi-eq}).
{}For the scalar mode, we use $\delta \phi$ as the representative one;
for theories without $\delta \phi$, like $f(R)$ gravity,
we can use (\ref{hat-phi-phi-pert}).

{}For the gravitational wave, from (\ref{MSF-GW-eq}-\ref{MSF-GW-SS-sol}) 
we have:
\bea
   & &  \ddot H_T + \left( 3 H + {\dot F \over F} \right) \dot H_T 
        - {1 \over a^2} \nabla^2 H_T = 0,
   \label{GGT-GW-eq} \\
   & & H_T ({\bf x}, t) = C_g ({\bf x})
       - D_g ({\bf x}) \int^t_0 {1\over a^3 F} dt,
   \label{GGT-GW-LS-sol} \\
   & & H_T ({\bf k}, \eta) = - { 1 \over a \sqrt{F} }
       \Big[ c_{g1} ({\bf k}) e^{ik\eta}
       + c_{g2} ({\bf k}) e^{-ik\eta} \Big].
   \label{GGT-GW-SS-sol}
\eea

In the Einstein gravity limit with a minimally coupled scalar field
we have $F = 1 = \omega$ and, apparently,
(\ref{BG-GGT1}-\ref{GGT-GW-SS-sol}) reduce to the corresponding 
equations in \S \ref{sec:MSF}.
Remarkably, the growing modes of $\delta \phi_\varphi$ and $H_T$ in
(\ref{GGT-delta-phi-LS-sol},\ref{GGT-GW-LS-sol}) do not involve
$F$ or $\omega$.
This implies that, for the growing mode in the large scale limit, 
the same solutions in the Einstein's gravity remain valid in the
generalized gravity theories.

Thus, we complete our derivation of the results valid in a class of
generalized Einstein's gravity theories directly from the known results
in the Einstein's gravity using the conformal transformation.
The method is generally valid independently of the gauge conditions.

%%%%%%%%%%%%%%%%%%%%%%%%%%%%%%%%%%%%%%%%%%%%%%%%%%%%%%%%%%%%%%%
\section{Comparison with the zero-shear gauge}
         \label{sec:ZSG}

The large scale asymptotic solutions in the zero-shear gauge
are derived in \cite{H-GGT1}-\cite{H-GGT3};
studies in \cite{Mukhanov-etal,Russian} are also based on this gauge condition.
{}From (16,18) of \cite{H-GGT3} we have:
\bea
   & & \delta \phi_\chi ({\bf x}, t) = - C ({\bf x})
       {\dot \phi \over a F} \int_0^t a F dt
       + d ({\bf x}) {\dot \phi \over a F},
   \label{ZSG-delta-phi} \\
   & & \varphi_\chi ({\bf x}, t) = C ({\bf x}) 
       \left( 1 - {H \over a F} \int_0^t a F dt \right) 
       + d ({\bf x}) {H \over a F},
   \label{ZSG-varphi}
\eea
where $d({\bf x})$ is a coefficient of the decaying solution.
Einstein's gravity limits of (\ref{ZSG-varphi},\ref{ZSG-delta-phi}) 
and the relation through the conformal equivalence can be shown easily;
to show that the following relations are useful:
\bea
   & & {\delta \hat \phi_\chi \over \dot {\hat \phi} }
       = {\delta \phi_\chi \over \dot \phi}
       = {\delta F_\chi \over \dot F}, \quad
       \hat \varphi_\chi = \varphi_\chi + {\delta F_\chi \over 2 F}.
   \label{ZSG-CT-relations}
\eea
{}From \S 2.2 of \cite{H-PRW} we note that
$\varphi_\chi \equiv \varphi - H \chi \equiv - H \chi_\varphi$ and
$\delta \phi_\chi \equiv \delta \phi - \dot \phi \chi$
are gauge invariant combination which becomes
$\varphi$ and $\delta \phi$, respectively, under the zero-shear gauge
which takes $\chi \equiv 0$ as the gauge condition.
Thus, from (\ref{UCG-UFG},\ref{ZSG-delta-phi},\ref{ZSG-varphi}) we 
can derive the solution in the large scale limit as
\bea
   & & \delta \phi_\varphi = \delta \phi - {\dot \phi \over H} \varphi
       = \delta \phi_\chi - {\dot \phi \over H} \varphi_\chi
       = - {\dot \phi \over H} C ({\bf x}).
   \label{UCG-ZSG}
\eea
Notice that the decaying solution in the zero-shear gauge
cancels out in the uniform-curvature gauge.
In fact, in \cite{H-GGT} we derived 
\bea
   & & D ({\bf x}) = - 2 \nabla^2 d ({\bf x}).
   \label{D-d}
\eea
Using the conformal transformation the proof of (\ref{D-d}) becomes simple:
In the case of minimally coupled scalar field, from (8-10) of \cite{H-MSF}
we can derive
\bea
   & & {1 \over a^2} \nabla^2 \varphi_\chi = {\dot H \over H}
       \dot \varphi_{\delta \phi}.
   \label{ZSG-UFG}
\eea
{}From (\ref{UCG-UFG},\ref{MSF-delta-phi-LS-sol},\ref{ZSG-varphi}) 
we can show (\ref{D-d}).
Since (\ref{D-d}) is invariant under the conformal transformation,
it remains valid for the generalized Einstein's gravity theories, 
thus is proved.
Thus, although it looks complicated, the decaying mode in the 
uniform-curvature gauge is {\it higher order} term in the 
large scale expansion compared with the one in the zero-shear gauge.
We also note that the growing mode solution is 
{\it simpler} in the uniform-curvature gauge; 
for example, it does not involve $F$ which characterizes the 
non-Einsteinian nature of the theory.
Furthermore, the small scale solution in the uniform-curvature gauge
(\ref{GGT-delta-phi-SS-sol}) is also much simpler than the one 
in the zero-shear gauge which is presented in \cite{H-GGT2}.
The equation for $\delta \phi_\chi$ in the minimally coupled scalar field
is presented in (37) of \cite{H-MSF} which can be compared with the
one for $\delta \phi_\varphi$ in (\ref{MSF-delta-phi-eq}).

%%%%%%%%%%%%%%%%%%%%%%%%%%%%%%%%%%%%%%%%%%%%%%%%%%%%%%%%%%%%%%%
\section{Quantization}
         \label{sec:Quantization}

In this section we present the normalization condition arising from the
canonical quantization of the perturbed part of the metric and the scalar field.
Instead of the classical decomposition we replace $\delta \phi ({\bf x},t)$ 
with a quantum (Heisenberg representation) operator
$\delta \hat \phi ({\bf x},t)$ as
\bea
   & & \phi ({\bf x}, t) = \bar \phi (t) + \delta \hat \phi ({\bf x}, t).
   \label{decomposition-quantum}
\eea
In this section an overhat indicates the quantum operator.
The corresponding perturbed parts of the metric and matter variables 
are similarly changed into operators.
We call this approach the {\it perturbative semiclassical approximation}
\cite{H-UCG}.
Since we are considering a {\it flat} three-space background we may expand
$\delta \hat \phi ({\bf x},t)$ in the following mode expansion
\bea
   & & \delta \hat \phi ( {\bf x}, t) = \int {d^3 k \over ( 2 \pi)^{3/2} }
       \left[ \hat a_{\bf k} \delta \phi_{\bf k} (t) e^{i {\bf k}\cdot {\bf x}}
       + \hat a^\dagger _{\bf k} \delta \phi^*_{\bf k} (t) e^{-i {\bf k} 
       \cdot {\bf x}} \right].
   \label{mode-expansion}
\eea
The annihilation and creation operators $\hat a_{\bf k}$ and 
$\hat a_{\bf k}^\dagger$ satisfy the standard commutation relations:
\bea
   & & [ \hat a_{\bf k} , \hat a_{{\bf k}^\prime} ] = 0, \quad
       [ \hat a^\dagger_{\bf k} , \hat a^\dagger_{{\bf k}^\prime} ] = 0, \quad
       [ \hat a_{\bf k} , \hat a^\dagger_{{\bf k}^\prime} ]
       = \delta^3 ( {\bf k} - {\bf k}^\prime ).
   \label{annihilation}
\eea
$\delta \phi_{\bf k} (t)$ is the mode function, a complex solution of
the classical mode evolution equation.
(\ref{GGT-delta-phi-eq}) becomes equation for $\delta \hat \phi_\varphi$.
Using (\ref{mode-expansion}) we can derive the equation for the mode function
$\delta \phi_{\varphi {\bf k}}$ which satisfies the same form as 
(\ref{GGT-delta-phi-eq}).

From the action for the perturbed scalar field in (\ref{Action-GGT}) 
we can derive the conjugate momenta and the commutation relation.
Using $S = \int {\cal L} dt d^3 x$ and $g^{(3)} = 1$, we have
\bea
   & & \delta \pi_\varphi ({\bf x}, t) 
       \equiv { \partial {\cal L} \over \partial \delta \dot \phi_\varphi }
       = a^3 { \omega + {3 \dot F^2 \over 2 \dot \phi^2 F}
       \over \left( 1 + {\dot F \over 2 H F} \right)^2 } 
       \delta \dot \phi_\varphi ({\bf x}, t).
   \label{conjugate-momentum}
\eea
Thus, from 
$[\delta \hat \phi_\varphi ({\bf x},t), 
\delta \hat \pi_\varphi ({\bf x}^\prime, t) ]
= i \delta^3 ({\bf x} - {\bf x}^\prime)$ 
we have the equal-time commutation relation
\bea
   & & [ \delta \hat \phi_\varphi ({\bf x},t), 
       \delta {\dot {\hat \phi}_\varphi} ({\bf x}^\prime, t) ] 
       = {i \over a^3} 
       { \left( 1 + {\dot F \over 2 H F} \right)^2 
       \over \omega + {3 \dot F^2 \over 2 \dot \phi^2 F} } 
       \delta^3 ({\bf x} - {\bf x}^\prime).
   \label{commutation-relation}
\eea
In order for (\ref{annihilation}) to be in accord with
(\ref{commutation-relation}), the mode function $\delta \phi_{\bf k} (t)$
should follow the Wronskian condition
\bea
   & & \delta \phi_{\varphi {\bf k}} \delta \dot \phi_{\varphi {\bf k}}^{*}
       - \delta \phi^*_{\varphi {\bf k}} \delta \dot \phi_{\varphi {\bf k}} 
       = {i \over a^3}
       { \left( 1 + {\dot F \over 2 H F} \right)^2
       \over \omega + {3 \dot F^2 \over 2 \dot \phi^2 F} }.
   \label{Wronskian}
\eea
(\ref{Wronskian}) can be used as the normalization condition
for the quantum fluctuation which fixes the amplitude up to vacuum choice;
the choice of vacuum depends on that of the mode function.

Definitions of the two-point function and the power spectrum 
based on the vacuum expectation value are presented in \S IIIA of 
\cite{H-QFT}.
In some special expansion stages in the case of the minimally coupled scalar 
field the rigorous analyses of the generated quantum fluctuations
are presented in \cite{H-QFT,H-UCG,H-IF}.

%%%%%%%%%%%%%%%%%%%%%%%%%%%%%%%%%%%%%%%%%%%%%%%%%%%%%%%%%%%%%%%
\section{Discussions}
         \label{sec:Discussion}

In this paper we have derived the equations and the general asymptotic 
solutions, characterizing the perturbed universe model, 
which are valid in a wide class of generalized gravity theories.
Straightforward derivations of the results in \S \ref{sec:GGT} 
without addressing to the conformal equivalence with the minimally coupled
scalar field are presented in \cite{H-GGT}.
The equations and the asymptotic solutions for the scalar and tensor modes 
in \S \ref{sec:GGT} can be written in unified forms; see Table 3.

\centerline{\bf LOCATE TABLE 3.}

\noindent
We note that growing modes of $\varphi_{\delta \phi}$ and $H_T$ are 
conserved in the large scale limit.
Ignoring the transient modes, from 
(\ref{UCG-UFG},\ref{GGT-delta-phi-LS-sol},\ref{GGT-GW-LS-sol}) we have:
\bea
   & & \varphi_{\delta \phi} ({\bf x}, t) = C ({\bf x}), \quad
       H_T ({\bf x}, t) = C_g ({\bf x}).
\eea

The generalized gravity theories we have considered in this paper
do not include the following types of generalized gravity theories
in the Lagrangian:
general couplings with the Ricci or the conformal curvature like
$R_{ab} R^{ab}$ or $C_{abcd} C^{abcd}$,
couplings involving general derivatives of the scalar curvature
and the scalar field, etc. 
Cosmological perturbation analyses in the context of these more generalized
gravity theories, and quantum field theoretical counterparts
of the classical results presented in this paper
are the challenging subjects left for future studies.

%%%%%%%%%%%%%%%%%%%%%%%%%%%%%%%%%%%%%%%%%%%%%%%%%%%%%%%%%%%%%%%
\acknowledgments

We thank Dr. H. Noh for valuable discussions.
This work was supported in part by the Korea Science and Engineering
Foundation through the SRC program of SNU-CTP.

%%%%%%%%%%%%%%%%%%%%%%%%%%%%%%%%%%%%%%%%%%%%%%%%%%%%%%%%%%%%%%%

\vskip 2cm
%%%%%%%%%%%%%%%%%%%%%%%%%%%%%%%%%%%%%%%%%%%%%%%%%%%%%%%%%%%%%%%
\baselineskip=13pt

\noindent
{\bf Table 1:} Cases of the generalized $f(\phi, R)$ gravity.

\noindent
-----------------------------------------------------------------------------------------------------------------------------------------------
\begin{tabbing}
$f(\phi,R)$ gravity \hskip 2.1cm
  \= $ L = {1\over 2} f(\phi, R)
       - {1 \over 2} \omega(\phi) \phi^{;a} \phi_{,a} - V(\phi) $ \hskip 1.1cm
  \= $ F = F(\phi,R)$
  \\
-----------------------------------------------------------------------------------------------------------------------------------------------
  \\
$f(R)$ gravity 
  \> $ L = {1\over 2} f(R) $ 
  \> $ F = F(R), \quad \phi = 0$
  \\
  \>
  \>
  \\
$R^2$ gravity
  \> $ L = {1 \over 2} \left( R - {R^2 \over 6 M^2} \right) $
  \> $ F = 1 - {R \over 3 M^2}, \quad \phi = 0$
  \\
  \>
  \>
  \\
Generalizes scalar
  \> $ L = \phi R - \omega (\phi) {\phi^{;a} \phi_{,a} \over \phi} - V (\phi) $
  \> $ F = 2 \phi, \quad \omega \rightarrow 2 {\omega(\phi) \over \phi}$
  \\
  \qquad tensor theory
  \>
  \>
  \\
Brans-Dicke theory
  \> $ L = \phi R - \omega {\phi^{;a} \phi_{,a} \over \phi} $
  \> $ F = 2 \phi, \quad \omega \rightarrow 2 {\omega \over \phi}, \quad
       V = 0 $
  \\
  \>
  \>
  \\
$F(\phi)R$ gravity
  \> $ L = {1 \over 2} F (\phi) R
       - {1\over 2} \omega (\phi) \phi^{;a} \phi_{,a} - V(\phi) $
  \> $ F = F(\phi)$
  \\
  \>
  \>
  \\
Dilaton gravity
  \> $ L = {1 \over 2} e^{-\phi} \left( R - \phi^{;a} \phi_{,a} \right) $
  \> $ F = e^{-\phi}, \quad \omega = e^{-\phi}, \quad V = 0$
  \\
  \>
  \>
  \\
Generally coupled 
  \> $ L = {1 \over 2} \left( \gamma - \xi \phi^2 \right) R
       - {1\over 2} \phi^{;a} \phi_{,a} - V (\phi) $
  \> $ F = \gamma - \xi \phi^2, \quad \omega = 1$
  \\
  \qquad scalar field
  \>
  \>
  \\
Nonminimally coupled
  \> $ L = {1 \over 2} \left( 1 - \xi \phi^2 \right) R
       - {1\over 2} \phi^{;a} \phi_{,a} - V (\phi) $
  \> $ F = 1 - \xi \phi^2, \quad \omega = 1$
  \\
  \qquad scalar field
  \>
  \>
  \\
Conformal coupling
  \> $ L = {1 \over 2} \left( 1 - {1 \over 6} \phi^2 \right) R
       - {1\over 2} \phi^{;a} \phi_{,a} - V (\phi) $
  \> $ F = 1 - {1\over 6} \phi^2, \quad \omega = 1$
  \\
  \>
  \>
  \\
Minimally coupled
  \> $ L = {1 \over 2} R - {1\over 2} \phi^{;a} \phi_{,a} - V (\phi) $
  \> $ F = 1, \quad \omega = 1$
  \\
  \qquad scalar field
  \>
  \>
  \\
Induced gravity
  \> $ L = {1 \over 2} \epsilon \phi^2 R - {1\over 2} \phi^{;a} \phi_{,a} 
       - V (\phi) $
  \> $ F = \epsilon \phi^2, \quad \omega = 1$
  \\
-----------------------------------------------------------------------------------------------------------------------------------------------
  \\
\end{tabbing}

\newpage
%%%%%%%%%%%%%%%%%%%%%%%%%%%%%%%%%%%%%%%%%%%%%%%%%%%%%%%%%%%%%%%
\noindent
{\bf Table 2:} Conformally transformed scalar field.  

\noindent
-----------------------------------------------------------------------------------------------------------------------------------------------
\begin{tabbing}
General forms \hskip 1.4cm
  \= $ d \hat \phi = \sqrt{{\omega \over F} d \phi^2 
       + {3 d F^2 \over 2 F^2} } $ \hskip 1.0cm
  \= $ \dot {\hat \phi} = \sqrt{ {\omega \over F} \dot \phi^2
       + {3 \dot F^2 \over 2 F^2} } $ \hskip 0.7cm
  \= $ {\delta \hat \phi \over \dot {\hat \phi} }
       = {\delta \phi \over \dot \phi} = {\delta F \over \dot F} $ \hskip .5cm
  \= $ \hat V = {RF - f + 2 V \over 2 F^2} $
  \\
-----------------------------------------------------------------------------------------------------------------------------------------------
  \\
$f(R)$ gravity, 
  \> $ \hat \phi = \sqrt{3 \over 2} \ln{F} $ 
  \> $ \hat \phi = \sqrt{3 \over 2} \ln{F} $ 
  \> $ \delta \hat \phi = \sqrt{3 \over 2} {\delta F \over F} $
  \> $ \hat V = {RF - f \over 2 F^2} $
  \\
$R^2$ gravity
  \>
  \>
  \>
  \>
  \\
Generalizes scalar
  \> $ \hat \phi = \int \sqrt{ \omega (\phi) + {3 \over 2} } 
       {d \phi \over \phi} $
  \> $ \dot {\hat \phi} = \sqrt{ \omega (\phi) + {3 \over 2} } 
       {\dot \phi \over \phi} $
  \> $ {\delta \hat \phi \over \dot {\hat \phi} }
       = {\delta \phi \over \dot \phi} $
  \> $ \hat V = {V \over 4 \phi^2} $
  \\
  \qquad tensor theory
  \>
  \>
  \>
  \\
Brans-Dicke theory
  \> $ \hat \phi = \sqrt{ \omega + {3 \over 2} } \ln{\phi} $
  \> $ \hat \phi = \sqrt{ \omega + {3 \over 2} } \ln{\phi} $
  \> $ \delta \hat \phi = \sqrt{ \omega + {3 \over 2} } 
       { \delta \phi \over \phi } $
  \> $ \hat V = 0 $
  \\
  \>
  \>
  \>
  \\
$F(\phi)R$ gravity
  \> $ \hat \phi = \int \sqrt{ {\omega \over F} 
       + {3 \over 2 F^2} \left( {dF \over d \phi} \right)^2 } d \phi $
  \> $ \hat \phi = \int \sqrt{ {\omega \over F} 
       + {3 \dot F^2 \over 2 F^2 \dot \phi^2} } d \phi $
  \> $ {\delta \hat \phi \over \dot {\hat \phi} }
       = {\delta \phi \over \dot \phi} $
  \> $ \hat V = {V \over 2 F^2} $
  \\
  \>
  \>
  \>
  \\
Dilaton gravity
  \> $ \hat \phi = \sqrt{5 \over 2} \phi $
  \> $ \hat \phi = \sqrt{5 \over 2} \phi $
  \> $ \delta \hat \phi = \sqrt{5 \over 2} \delta \phi $
  \> $ \hat V = 0 $
  \\
  \>
  \>
  \>
  \\
Generally coupled 
  \> $ \hat \phi = \int { \sqrt{ \gamma + \xi ( 6 \xi -1 ) \phi^2 }
       \over \gamma - \xi \phi^2 } d \phi $
  \> $ \dot {\hat \phi} = { \sqrt{ \gamma + \xi ( 6 \xi -1 ) \phi^2 }
       \over \gamma - \xi \phi^2 } \dot \phi $
  \> $ {\delta \hat \phi \over \dot {\hat \phi} }
       = {\delta \phi \over \dot \phi} $
  \> $ \hat V = {V \over \gamma - \xi \phi^2} $
  \\
  \qquad scalar field
  \>
  \>
  \>
  \\
Nonminimally coupled
  \> $ \hat \phi = \int { \sqrt{ 1 + \xi ( 6 \xi -1 ) \phi^2 }
       \over 1 - \xi \phi^2 } d \phi $
  \> $ \dot {\hat \phi} = { \sqrt{ 1 + \xi ( 6 \xi -1 ) \phi^2 }
       \over 1 - \xi \phi^2 } \dot \phi $
  \> $ {\delta \hat \phi \over \dot {\hat \phi} }
       = {\delta \phi \over \dot \phi} $
  \> $ \hat V = {V \over 1 - \xi \phi^2} $
  \\
  \qquad scalar field
  \>
  \>
  \>
  \\
Conformal coupling
  \> $ \hat \phi = \sqrt{6} \tanh^{-1}{\phi \over \sqrt{6}} $
  \> $ \hat \phi = \sqrt{6} \tanh^{-1}{\phi \over \sqrt{6}} $
  \> $ \delta \hat \phi = {\delta \phi \over 1 - {1\over 6} \phi^2 } $
  \> $ \hat V = {V \over 1 - {1 \over 6} \phi^2} $
  \\
  \>
  \>
  \>
  \\
Minimally coupled
  \> $ \hat \phi = \phi$
  \> $ \hat \phi = \phi$
  \> $ \delta \hat \phi = \delta \phi$
  \> $ \hat V = V $
  \\
  \qquad scalar field
  \>
  \>
  \>
  \\
Induced gravity
  \> $ \hat \phi = \sqrt{ 6 + {1\over \epsilon} } \ln{\phi} $
  \> $ \hat \phi = \sqrt{ 6 + {1\over \epsilon} } \ln{\phi} $
  \> $ \delta \hat \phi = \sqrt{ 6 + {1\over \epsilon} } 
       {\delta \phi \over \phi} $
  \> $ \hat V = {V \over \epsilon \phi^2} $
  \\
-----------------------------------------------------------------------------------------------------------------------------------------------
  \\
\end{tabbing}

%%%%%%%%%%%%%%%%%%%%%%%%%%%%%%%%%%%%%%%%%%%%%%%%%%%%%%%%%%%%%%%
\noindent
{\bf Table 3:} Unified form expressions for the scalar and tensor modes.  
               We have $v \equiv - z \Phi$ and $z \equiv a \sqrt{Q}$.

\noindent
-----------------------------------------------------------------------------------------------------------------------------------------------
\begin{tabbing}
Action (for \hskip 1cm
  \= $ S = {1 \over 2} \int a^3 Q \left( \dot \Phi^2 
       - {1 \over a^2} \Phi^{|\alpha} \Phi_{,\alpha} \right)
       \sqrt{ g^{(3)} } dt d^3 x $ \hskip 0.6cm
  \= $ S = {1 \over 2} \int \left( v^{\prime 2} - v^{|\alpha} v_{,\alpha} 
       + {z^{\prime \prime} \over z} v^2 \right)
       \sqrt{ g^{(3)} } d \eta d^3 x $
  \\
  \quad scalar mode)
  \\
Equation 
  \> $ \ddot \Phi + \left( 3 H + {\dot Q \over Q} \right) \dot \Phi
       - {1 \over a^2} \nabla^{2} \Phi = 0 $ 
  \> $ v^{\prime\prime} - \left( \nabla^2 + {z^{\prime\prime} \over z}
       \right) v = 0 $
  \\
  \>
  \\
Large scale
  \> $ \Phi = C ({\bf x}) - D ({\bf x}) \int_0^t (a^3 Q)^{-1} dt $
  \> $ v = - z \left[ C ({\bf x}) - D ({\bf x}) \int_0^\eta 
       z^{-2} d \eta \right] $
  \\
  \>
  \\
Small scale
  \> $ \Phi = - (a \sqrt{Q})^{-1}
       \left[ c_1 ({\bf k}) e^{ik \eta} + c_2 ({\bf k}) e^{-ik \eta} \right] $
  \> $ v = c_1 ({\bf k}) e^{ik \eta} + c_2 ({\bf k}) e^{-ik \eta} $
  \\
-----------------------------------------------------------------------------------------------------------------------------------------------
  \\
Scalar mode
  \> $ \Phi = \varphi_{\delta \phi}, \quad 
       Q = { \omega \dot \phi^2 + {3 \dot F^2 \over 2 F} 
       \over \left( H + {\dot F \over 2 F} \right)^2 } $
  \> $ v = - z \varphi_{\delta \phi}, \quad z = a 
       { \sqrt{ \omega \dot \phi^2 + {3 \dot F^2 \over 2 F} }
       \over H + {\dot F \over 2 F} } $
  \\
  \>
  \\
Tensor mode
  \> $ \Phi = H_T, \quad Q = F $
  \> $ v = - z H_T, \quad z = a \sqrt{F} $
  \\
-----------------------------------------------------------------------------------------------------------------------------------------------
  \\
\end{tabbing}

%%%%%%%%%%%%%%%%%%%%%%%%%%%%%%%%%%%%%%%%%%%%%%%%%%%%%%%%%%%%%%%

\begin{references}
\bibitem{H-GGT}
         J. Hwang, Phys. Rev. D {\bf 53}, 762 (1996);
         J. Hwang and H. Noh, unpublished (1996).
\bibitem{H-GGT1}
         J. Hwang, Class. Quantum Grav. {\bf 7}, 1613 (1990).
\bibitem{H-GGT2}
         J. Hwang, Phys. Rev. D {\bf 42}, 2601 (1990).
\bibitem{H-GGT3}
         J. Hwang, Class. Quantum Grav. {\bf 8}, 195 (1991).
\bibitem{H-QFT}
         J. Hwang, Phys. Rev. D {\bf 48}, 3544 (1993).
\bibitem{H-UCG}
         J. Hwang, Class. Quantum Grav. {\bf 11}, 2305 (1994).
\bibitem{H-MSF}
         J. Hwang, Astrophys. J. {\bf 427}, 542 (1994).
\bibitem{H-IF}
         J. Hwang, J. Korean Phys. Soc. {\bf 28}, S502 (1995);
         J. Hwang and H. Minn, in Proceedings of the second Alexander
            Friedmann international seminar on gravitation and cosmology,
            edited by Yu. N. Gnedin, A. A. Grib and V. M. Mostepanenko
            (St. Petersburg, Friedmann Laboratory Publishing, 1994), 265.
\bibitem{Dicke}
         R. H. Dicke, Phys. Rev. {\bf 125}, 2163 (1962);
         B. Whitt, Phys. Lett. B {\bf 145}, 176 (1984);
         K. Maeda, Phys. Rev. D {\bf 39}, 3159 (1989).
\bibitem{Brans}
         C. H. Brans, Class. Quant. Grav. {\bf 5}, L197 (1988);
         Y. M. Cho, Phys. Rev. Lett. {\bf 68}, 3133 (1992).
\bibitem{H-PRW}
         J. Hwang, Astrophys. J. {\bf 375}, 443 (1991).
\bibitem{Mukhanov-etal}
         V. F. Mukhanov, H. A. Feldman and R. H. Brandenberger, Phys. Rep.
               {\bf 215}, 203 (1992);
\bibitem{Russian}
         L. A. Kofman and V. F. Mukhanov, JETP Lett. {\bf 44}, 619 (1986);
         V. F. Mukhanov, L. A. Kofman and D. Yu. Pogosyan,
               Phys. Lett B {\bf 193}, 427 (1987);
         L. A. Kofman, V. F. Mukhanov and D. Yu. Pogosyan,
               Sov. Phys. JETP {\bf 66}, 433 (1988);
         V. F. Mukhanov, Phys. Lett., B {\bf 218}, 17 (1989);
         M. B. Baibosunov, V. Ts. Gurovich and U. M. Imanaliev,
               Sov. Phys. JETP {\bf 71}, 636 (1990).
\end{references}
\end{document}